# The Geminid meteoroid stream as a potential meteorite dropper: a case study

José M. Madiedo[1, 2], Josep M. Trigo-Rodríguez[3], Alberto J. Castro-Tirado[4], José L. Ortiz[4] and Jesús Cabrera-Caño[2]

[1] Facultad de Ciencias Experimentales, Universidad de Huelva. 21071 Huelva (Spain).
[2] Departamento de Física Atómica, Molecular y Nuclear. Facultad de Física. Universidad de Sevilla. 41012 Sevilla, Spain.
[3] Institute of Space Sciences (CSIC-IEEC), Campus UAB, Facultat de Ciències, Torre C5-parell-2ª, 08193 Bellaterra, Barcelona, Spain.
[4] Instituto de Astrofísica de Andalucía, CSIC, Apt. 3004, Camino Bajo de Huetor 50, 18080 Granada, Spain.

**ABSTRACT**
A Geminid fireball with an absolute magnitude of -13 was observed over the south of Spain on Dec. 15, 2009. This extraordinarily bright event (the brightest Geminid ever recorded by our team) was imaged from two meteor observing stations operated by the SPanish Meteor Network (SPMN). The bolide exhibited fast and quasi-periodic variations in brightness, a behaviour typically associated to the rotation of the parent meteoroid. The inferred tensile strength of this particle was found to be significantly higher that the typical values obtained for Geminid meteoroids. The fireball penetrated in the atmosphere till a final height of about 25 km above the ground level and a non-zero terminal mass was calculated at the ending point of the luminous trajectory. In this way, the observational evidence points to the existence of a population of meteoroids in the higher end of the Geminid mass distribution capable of producing meteorites. From the relative chemical abundances inferred from the emission spectrum of this bolide we conclude that the Geminid-forming materials are similar to some primitive carbonaceous chondrite groups. Then, we conclude that in meteorite collections from cold deserts, capable to preserve meteorites of few tens of grams, some rare groups of carbonaceous chondrites could be coming from the Geminid parent body: (3200) Phaeton.

## 1 INTRODUCTION

The Geminids is the densest annual meteoroid stream and its structure and plausible origin have been extensively studied (Fox et al. 1983; Ryabova 2007, 2008, 2012; Webster et al. 1966; Williams and Wu 1993; Williams and Ryabova 2011). Thus, opposite formation models have been proposed, invoking either the fragmentation of an asteroid or a cometary-driven origin (Jones and Hawkes 1986; Gustafson 1989; Williams and Wu 1993). Its activity period extends from about November 27 to December 18, peaking around December 14 (Jenniskens 2006). Its generally accepted parent body is asteroid (3200) Phaethon, which is considered by some researchers as an extinct cometary nucleus rather than a regular asteroid (Čapek & Borovička 2009). However, on the basis of spectral and dynamical similarities it has been recently proposed that asteroid 2 Pallas is the likely parent body of asteroid (3200) Phaeton (De León et al. 2010). Then, in such scenario (3200) Phaeton should be a fragment of 2 Pallas produced as a by-product of a disintegration cascade, and evolving inwards to regions where perhaps the body was disrupted producing the Geminid stream. The perihelion of this stream is of about 0.14 AU, and this relatively small value has been correlated to the observed depletion of volatiles in Geminid meteoroids (Kasuga et al. 2006; Čapek & Borovička 2009).

One of the aims of the SPMN is the analysis of very bright bolides, as these may give rise to meteorite falls. Nevertheless, meteorite-dropper fireballs are rare events, as even meter sized cometary meteoroids are completely destroyed at high altitudes in the atmosphere (Madiedo et al. 2013d). However, the asteroidal origin of the Geminids suggests that, under appropriate conditions, this stream could be a potential meteorite producer. A very bright Geminid fireball recorded over Spain on December 15, 2009 supports this idea. We analyze here this extraordinary event, which in addition exhibited the phenomenon known as "meteor flickering", a behaviour related to the rotation of the parent meteoroid. Besides, as a result of the systematic spectroscopic monitoring developed at one of the meteor observing stations that recorded this fireball, the emission spectrum produced during the ablation of the meteoroid in the atmosphere was also obtained. The analysis of this spectrum has provided the relative abundances of the main rock-forming elements in this meteoroid.

## 2 INSTRUMENTATION AND METHODS

The Geminid fireball analyzed here was simultaneously imaged from two SPMN meteor observing stations: Sevilla and El Arenosillo. Their geographical coordinates are given in Table 1. Station #1 employed an array of low-lux CCD video cameras manufactured by Watec Corporation (models 902H2 and 902H2 Ultimate). These devices, that generate imagery according to the PAL video standard, work in an autonomous way by means of the MetControl software (Madiedo et al. 2010). The field of view covered by each CCD video camera ranges, approximately, from 62x50 to 14x11 degrees. A more detailed description about the operation of these systems, which can image meteors with brightness higher than magnitude +3/+4, is given elsewhere (Trigo-Rodriguez et al. 2007, 2008, 2009; Madiedo & Trigo-Rodríguez 2007, 2010). Some of these CCD devices are configured as video spectrographs by attaching holographic diffraction gratings (500 or 1000 lines/mm, depending on the device) to the objective lens. These spectrographs can image spectra for meteor events with brightness higher than mag. -3/-4. In this way, we can infer information about the chemical composition of meteoroids ablating in the atmosphere (Trigo-Rodriguez et al. 2009; Madiedo et al. 2013a, 2013b, 2013c).

On the other hand, station #2 operates an automated all-sky low-scan-rate CCD camera named CASANDRA-1 (Trigo-Rodríguez et al. 2004b). This device, which was designed and developed by Castro-Tirado et al. (2008), is endowed with a monochrome high-resolution (4096x4096 pixels) CCD sensor cooled to $\Delta T = 40$ ºC below ambient temperature and records images of the night sky with 60 s exposures. It can image meteors trails with brightness higher than magnitude +2/+3 (Trigo-Rodríguez et al. 2008).

For data reduction we have followed the technique described in Trigo-Rodriguez et al. (2007) and Madiedo et al. (2013a, 2013b). Thus, meteor positions were measured by hand in both, the image taken by the slow-scan-rate all-sky CCD camera and the video frames recorded by our CCD video devices. In this way we obtained the plate (x, y) coordinates of the fireball along its apparent path from each observing station. These values were then introduced in the AMALTHEA software (Trigo-Rodríguez et al. 2009; Madiedo et al. 2011a), which converted plate coordinates into equatorial coordinates by using the position of reference stars appearing in the images. This package employs the method of the intersection of planes to determine the position of the apparent radiant and also to reconstruct the trajectory in the atmosphere of the fireball (Ceplecha 1987). The orbital parameters of the parent meteoroid were also obtained by following the procedure described in Ceplecha (1987).

### 3 OBSERVATIONS

The New Moon and optimal weather conditions over the south-west of Spain on December 15, 2009 allowed us to monitor the activity of the Geminids under very favourable conditions. In this context, our CCD cameras recorded a very bright fireball at 04h20m11.9s±0.1s UTC (Figure 1), about two hours before local sunrise. Three CCD video cameras operating at Sevilla imaged its atmospheric path and two of them recorded the whole apparent trajectory, but the final part of the luminous path was out of the field of view of CASANDRA-1, the all-sky CCD device operating at El Arenosillo. In addition, one of our video spectrographs at Sevilla also imaged the emission spectrum produced during the ablation of the parent meteoroid in the atmosphere. This device also recorded the initial part of the luminous trajectory of the fireball. So, in total, four video recordings of the event were obtained from Sevilla. This bolide was named "Higuera de la Sierra", as the projection of the ground of the terminal point of its luminous path was located in the surroundings of this village in the province of Huelva. Its absolute magnitude was -13.0±0.5, making this the brightest Geminid ever recorded by our team. This bolide was recorded about one day after the peak activity of the Geminids in 2009, which took place on December 14 at around 05:10 UTC. Figure 1c shows the apparent trajectory of the fireball as seen from both meteor observing stations. The initial (preatmospheric) velocity of the event, and also the velocity along the meteor trajectory, were inferred from the video recordings provided by the CCD video cameras operating at Sevilla (station #1), since the all-sky CCD camera at El Arenosillo (station #2) is not endowed with a rotary shutter to insert time marks on the images. In order to minimize the errors in the determination of the preatmospheric velocity ($V_\infty$) and the apparent radiant position (which in turns minimizes the errors in the orbital parameters), the image taken by the high-resolution all-sky camera at station #2 was combined with the recordings from the CCD video camera at station #1 which provided the narrowest field of view (about 14x11 degrees) and recorded the beginning of the meteor trail. In addition, to get benefit from the high resolution provided by the all-sky camera at station #2 (4096x4096 pixels), just the initial portion of the luminous trajectory as observed from there (which was equivalent to a length of about 20 km along the atmospheric path of the fireball) was considered for this calculation. Thus, the part of the luminous path appearing near from the border of the image, where the event is spread over a large number of pixels, was not taken into consideration to obtain the apparent radiant and $V_\infty$. In this way, we obtained that the parent meteoroid impacted the atmosphere with a velocity $V_\infty$=35.7±0.5 km s$^{-1}$, with the apparent radiant located at the position α=116.72±0.08 °, δ=32.93±0.05 °. To determine the initial and final heights, the all-sky image taken from station #2 was combined with the recordings from the CCD cameras operating from Sevilla that imaged the beginning of the bolide and its terminal position. Thus, we inferred that the luminous phase started at a height of 89.1±0.5 km above the ground level and the inclination of the trajectory with respect the Earth's surface was of about 70º. The fireball penetrated the atmosphere till a height of 24.8±0.30.5 km. These trajectory and radiant parameters are summarized in Table 2. This information allowed us to determine the heliocentric orbit of the meteoroid (Table 3). The projection on the ecliptic plane of this orbit is plotted in Figure 1d. These data confirm the association of the "Higuera de la Sierra" fireball with the Geminid meteoroid stream.

**4 DISCUSSION**
**4.1. Photometric profile**

The photometric analysis of the images recorded for the SPMN151209 Geminid fireball provided the light curve shown in Figure 2. As can be seen, the fireball exhibited a sudden and brief flare at a time t=1.6 seconds after the beginning of the event. The flickering behaviour of the meteoroid can also be noticed and is more obvious just after the brightest phase. These rapid and quasi-periodic oscillations in luminosity have been previously found in Geminid fireballs (Beech et al. 2003; Babadzhanov & Konovalova 2004), but also in bolides associated to other meteor streams and events with a sporadic origin (Beech & Brown 2000; Beech 2001; Konovalova 2003). This light curve was employed to infer the preatmospheric mass of the meteoroid, $m_\infty$. Thus, this parameter was calculated by equating it to the total mass lost due to ablation between the beginning and the end of the luminous phase of the event:

$$m_\infty = 2 \int_{t_b}^{t_e} I/(\tau v^2) dt \qquad (1)$$

where $t_b$ and $t_e$ are, respectively, the times corresponding to the beginning and the end of the luminous phase. The luminous efficiency ($\tau$), which depends on velocity (v), was calculated by means of the relationships given by Ceplecha & McCrosky (1976). I is the measured luminosity of the fireball, which is related to the absolute magnitude M by means of

$$M = -2.5 \log (I) \qquad (2)$$

In this way, the mass of the meteoroid yields $m_\infty$=757±72 g. By assuming a bulk density $\rho_m$=2.9 g cm$^{-3}$ for Geminid meteoroids (Babadzhanov & Kokhirova 2009), the diameter of this particle yields 7.9±0.2 cm. One possibility to explain the observed flickering behaviour is that this body was non-rotating before its encounter with the Earth, but it gained rotation in the atmosphere. However, this mechanism is possible only for Geminid meteoroids with initial masses below $6·10^{-7}$ g (Babadzhanov & Konovalova 2004). So, we must conclude that the SPMN151209 meteoroid was rotating before impacting the atmosphere. The light curve reveals a flickering frequency of about 12 Hz, with a maximum flickering modulation amplitude of 0.4 magnitudes. By assuming a homogeneous ellipsoidal profile for the meteoroid, the flickering amplitude would vary as (Beech et al. 2003)

$$\Delta m = 2.5 \log(a/b) \qquad (3)$$

where a and b are, respectively, the semimajor and semi-minor axes of the body. Thus, to account for the observed maximum amplitude an axis ratio a/b=1.4 is required.

**4.2 Tensile strength**

We have considered that the main flare exhibited by the fireball (Figure 2) took place as a consequence of the sudden disruption of the meteoroid. This break-up, however, was not catastrophic, as the remaining material continued penetrating in the atmosphere till a final height of 24.8±0.5 km. The equation given by Bronshten (1981) provides the tensile (aerodynamic) strength S at which this fracture occurred:

$$S = \rho_{atm} \cdot v^2 \qquad (4)$$

Here v is the velocity of the body at the disruption point and $\rho_{atm}$ the atmospheric density at the corresponding height. This aerodynamic strength S can be used as an estimation of the tensile strength of the meteoroid (Trigo-Rodriguez & Llorca 2006, 2007). The air density at the disruption height (39±1 km above the ground level) yields $4.1 \cdot 10^{-6}$ g cm$^{-3}$. This value was estimated by means of the US standard atmosphere model (U.S. Standard Atmosphere 1976). At this stage, the velocity of the particle was of around 29.5 km s$^{-1}$ and, so, the aerodynamic strength yields $(3.8\pm0.4) \cdot 10^7$ dyn cm$^{-2}$. This value is higher than the typical aerodynamic strength exhibited by Geminid meteoroids (Trigo-Rodriguez & Llorca 2006, 2007), which supports the idea of a population of cm-sized tougher meteoroids in the Geminid stream. The existence of such population is supported by the analysis performed by Čapek & Vokrouhlický (2012) on the basis of the influence of thermal stress on Geminid meteoroids. This, in addition, would favour the possibility of a meteorite fall from members of this population.

**4.3 Dark flight and meteorite survival**

The deep penetration in the atmosphere exhibited by the SPMN151209 fireball and the high tensile strength of the meteoroid support the idea that this bolide was a potential meteorite dropper. So, we have estimated the mass surviving the ablation process by means of the following equation (Ceplecha et al. 1983):

$$m_E = \left( \frac{1.2 \rho_E v_E^2}{(dv/dt)_E \rho_m^{2/3}} \right) \qquad (5)$$

where $m_E$, $v_E$ and $(dv/dt)_E$ are, respectively, the mass, velocity and deceleration at the terminal point of the luminous trajectory of the fireball. $\rho_m$ is the meteoroid density and $\rho_E$ the air density at this terminal height. The latter was again calculated by using the US standard atmosphere model (U.S. Standard Atmosphere 1976).

From the analysis of the atmospheric trajectory we obtained $v_E$=7.5±1.5 km s$^{-1}$ and $(dv/dt)_E$= -27±3 km s$^{-2}$. So, the terminal mass yields $m_E$=25±19 g. The dark flight of the surviving mass was modelled by means of the AMALTHEA software, which performs this analysis by following the procedure described in (Ceplecha 1987). Atmospheric data provided by AEMET (the Spanish meteorological agency) were used to include wind effects in the calculations. We have assumed that the surviving mass corresponded to a single particle, which in addition was supposed to be spherical. According to this, the likely landing point would be located around the geographical coordinates 37.825°

N, 6.486º W, which corresponds to a countryside area in the north-east of the province of Huelva. However, the small terminal mass implies a very big uncertainty in the determination of the impact point. Besides, the type of vegetation that predominates in that area makes a search very unfavourable.

**4.4 Emission spectrum**
The reduction of the emission spectrum and the identification of the main emission lines appearing in it were performed in the usual way (Madiedo et al. 2013a, 2013b). So, after the video sequence recorded by the spectrograph was deinterlaced, the frames containing the spectrum were flat fielded and dark substracted. Typical emission lines produced by meteor plasmas were employed to calibrate the signal in wavelengths. The resulting curve, after correcting the spectrum by taking into account the spectral sensitivity of the device, is shown in Figure 3. Multiplet numbers are given after Moore (1945). As can be noticed, the emissions from the Na I-1 doublet at 588.9 nm and the Mg-I-2 triplet at 516.7 nm are very prominent, together with the contribution from the O I triplet in the infrared, at 777.4 nm. These, however, are not resolved in the spectrum. The intensity of the atmospheric $N_2$ bands in the red region is also very remarkable. Nevertheless, most of the lines identified in the spectrum correspond to Fe I, although these appear blended with other contributions.

The relative abundances of the main rock-forming elements in the meteoroid were obtained with a software application that reconstructs a synthetic spectrum by adjusting the temperature (T), the column density of atoms (N), the damping constant (D) and the surface area (P) from the observed brightness of lines, as explained in Trigo-Rodríguez et al. (2003). Then, the modelled spectrum of the meteor column is directly compared with the observed one to fit all its peculiarities. Once the above-mentioned physical parameters are fixed, the abundances of the main rocky elements can be modified in successive iterative steps until obtaining an optimal fit. The relative elemental abundances that we provide here are obtained as a consequence of such a fitting process. The elemental abundances derived by the software are relative to Fe. Consequently, the final computed values were multiplied by Fe/Si=1.16 (Anders and Grevesse 1989) to obtain the elemental abundances relative to Si. From the relative abundance ratios obtained, the materials released from 3200 Phaeton seem to be similar to some primitive carbonaceous chondrite groups like e.g. CI and CM (see Table 4). This result is similar to the obtained for a -5 magnitude photographic Geminid analysed in Trigo-Rodríguez et al. (2004a), but the fireball studied here has closer Mg and Na abundances. The Ca/Si remarkably low abundance deserves additional discussion as it is probably a lower limit. The reason is that Ca (like also Al) is typically underestimated in meteor spectra as a consequence of the inefficiency, particularly for low or mid-geocentric velocity meteoroids like the studied here, to bring all available Ca (or Al) to the vapour phase. Obviously we can obtain more reliable elemental ratios for these elements efficiently contributing to produce the emitted light. In fact, we previously noted (Trigo-Rodríguez et al. 2003) that both Ca and Al are associated with refractory minerals that are not fully vaporised during atmospheric ablation. This is why the contribution of these elements to the emitted light increases for meteoroids with higher geocentric velocity (Trigo-Rodríguez et al. 2003, 2004a; Trigo-Rodríguez and Llorca 2007).

Our conclusions can have important implications in Meteoritics because we have found clear evidence supporting that some meteoroids in the higher mass end of the Geminid stream can survive their atmospheric entry in the right geometric circumstances, having terminal masses of a few tens of grams. Besides, it has been established that small meteorites within this range of masses coming from Antarctica or other cold deserts are more efficiently preserved (Harvey 2003). Then, the possibility of having (3200) Phaeton (and 2 Pallas from our previous discussion) as source of meteorites to the Earth should be firmly considered. Then, we suggest that it is likely that in meteorite collections from Antarctica some rare chondrite group or grouplet represented by specimens with a mass of a few tens or hundreds of come from (3200) Phaeton.

## 5 CONCLUSIONS

We have analyzed a very bright Geminid fireball observed over the south-west of Spain on Dec. 15, 2009. The following conclusions were derived from this work:

1) The absolute magnitude of the event was $-13.0\pm0.5$. Its atmospheric trajectory was determined and the heliocentric orbit of the parent meteoroid was calculated. The bolide exhibited a flickering effect as a consequence of the rotation of the body. This particle had an initial mass of $757\pm72$ g and a diameter of about $7.9\pm0.2$ cm. In addition, this body exhibited a high tensile strength and penetrated the atmosphere till a final height of $24.8\pm0.5$ km above the ground level.
2) Although small, a non-zero terminal mass was calculated at the terminal point of the luminous phase of this deep-penetrating event. This supports the existence of tough and large enough meteoroids within the Geminid stream which are capable, under appropriate conditions, of producing meteorites.
3) The analysis of the emission spectrum produced by the fireball provided the relative abundances of the main rock-forming elements in the meteoroid. We have found that the SPMN151209 Geminid meteoroid had abundances close to the expected for primitive carbonaceous chondrite groups.
4) Consequently, we think that we have found clear evidence supporting that the Geminid stream is populated by meteoroids with enough mass to survive as meteorites with a few tens or hundreds of grams. Then it is likely that, preferentially in the current meteorite collections from Antarctica or other cold deserts where small meteorites are more efficiently preserved, some primitive chondrites could be free-delivered samples from (3200) Phaeton, and knowing its presumable origin, also from 2 Pallas.

**ACKNOWLEDGEMENTS**
We thank support from the Spanish Ministry of Science and Innovation (projects AYA2009-13227 and AYA 2009-14000-C03-01) and Junta de Andalucía (projects P09-FQM-4555 and P06-FQM-02192).

**TABLES**

Table 1. Geographical coordinates of the meteor observing stations involved in this work.

| Station # | Station name | Longitude (W) | Latitude (N) | Alt. (m) |
|---|---|---|---|---|
| 1 | Sevilla | 05º 58´ 50" | 37º 20´ 46" | 28 |
| 2 | El Arenosillo | 06º 43´ 58" | 37º 06´ 16" | 40 |

Table 2. Trajectory and radiant data for the SPMN151209 fireball (J2000). $H_b$ and $H_e$ are, respectively, the beginning and ending heights of the luminous phase. $H_m$ is the height corresponding to the maximum brightness. $\alpha_g$, $\delta_g$: right ascension and declination of the geocentric radiant. $V_\infty$, $V_g$, $V_h$: observed preatmospheric, geocentric and heliocentric velocities.

| $H_b$ (km) | $H_m$ (km) | $H_e$ (km) | $\alpha_g$ (º) | $\delta_g$ (º) | $V_\infty$ (km s$^{-1}$) | $V_g$ (km s$^{-1}$) | $V_h$ (km s$^{-1}$) |
|---|---|---|---|---|---|---|---|
| 89.1±0.5 | 39±1 | 24.8±0.5 | 114.85±0.09 | 32.37±0.06 | 35.7±0.5 | 34.1±0.5 | 33.7±0.5 |

Table 3. Orbital parameters (J2000) for the SPMN151209 fireball.

| a (AU) | e | i (º) | ω (º) | Ω (º) |
|---|---|---|---|---|
| 1.33±0.04 | 0.891±0.006 | 23.6±0.7 | 324.0±0.1 | 263.1891±10$^{-4}$ |

Table 4. Computed elemental abundances relative to Si of the SPMN151209 Geminid fireball compared with the inferred for other Solar System undifferentiated materials (Jessberger et al. 1988; Rietmeijer & Nuth 2000; Rietmeijer 2002). The GEMr spectrum was discussed in Trigo-Rodríguez et al. (2004a)

| | Mg | Na | Ca | Mn (×10$^{-4}$) | Ni (×10$^{-3}$) | T (K) |
|---|---|---|---|---|---|---|
| SPMN151209 | 1.10 | 0.06 | 0.030 | 80 | 30 | 4,500 |
| GEMr spectrum | 0.90 | 0.12 | 0.018 | 49 | - | 4,300 |
| 1P/Halley | 0.54 | 0.054 | 0.034 | 30 | 22 | - |
| IDPs | 0.85 | 0.085 | 0.048 | 150 | 37 | - |
| CI chondrites | 1.06 | 0.060 | 0.071 | 90 | 51 | - |
| CM chondrites | 1.04 | 0.035 | 0.072 | 60 | 46 | - |

**FIGURES**

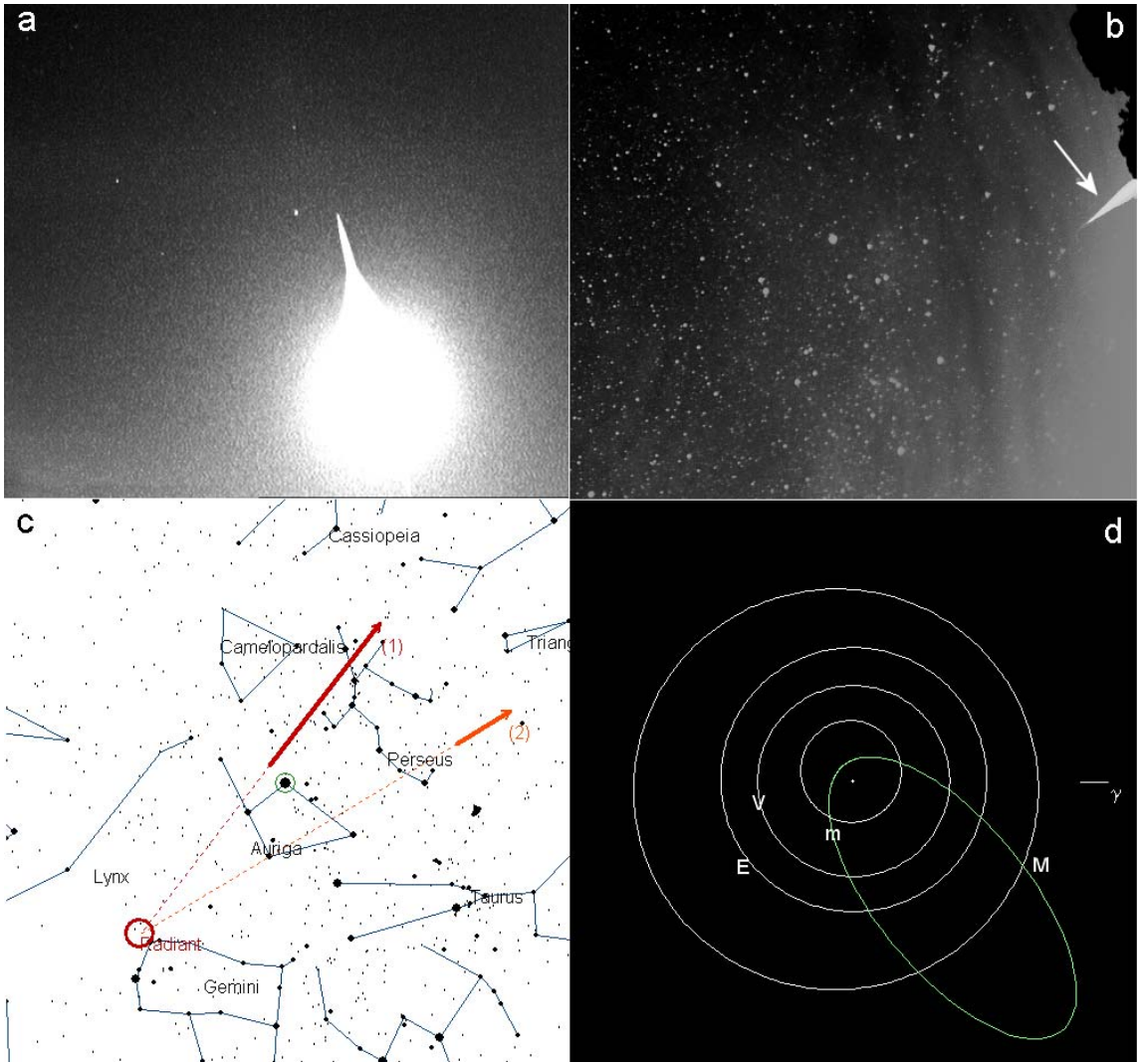

Figure 1. Composite images of the "Higuera de la Sierra" fireball (SPMN151209) recorded from Sevilla (a) and El Arenosillo (b). c) Apparent trajectory in the sky as observed from (1) Sevilla and (2) El Arenosillo. d) Projection on the ecliptic plane of the heliocentric orbit of the parent meteoroid. The orbits of Mercury (m), Venus (V), Earth (E) and Mars (M) are included for comparison.

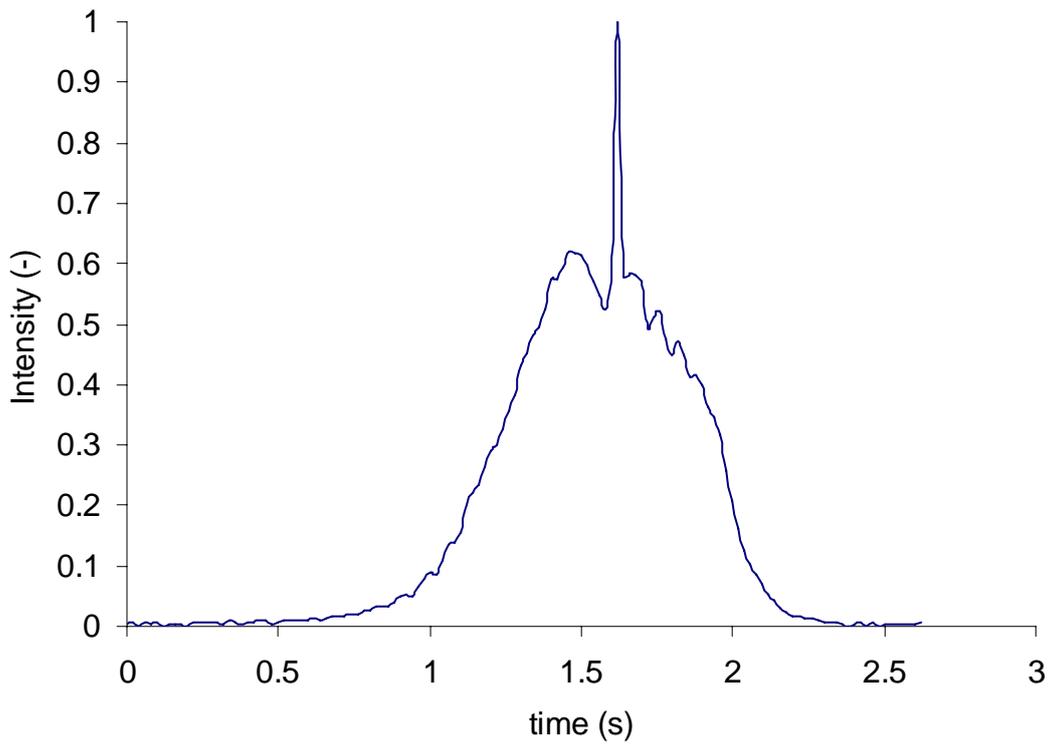

Figure 2. Light curve (relative pixel intensity vs. time) of the SPMN151209 fireball.

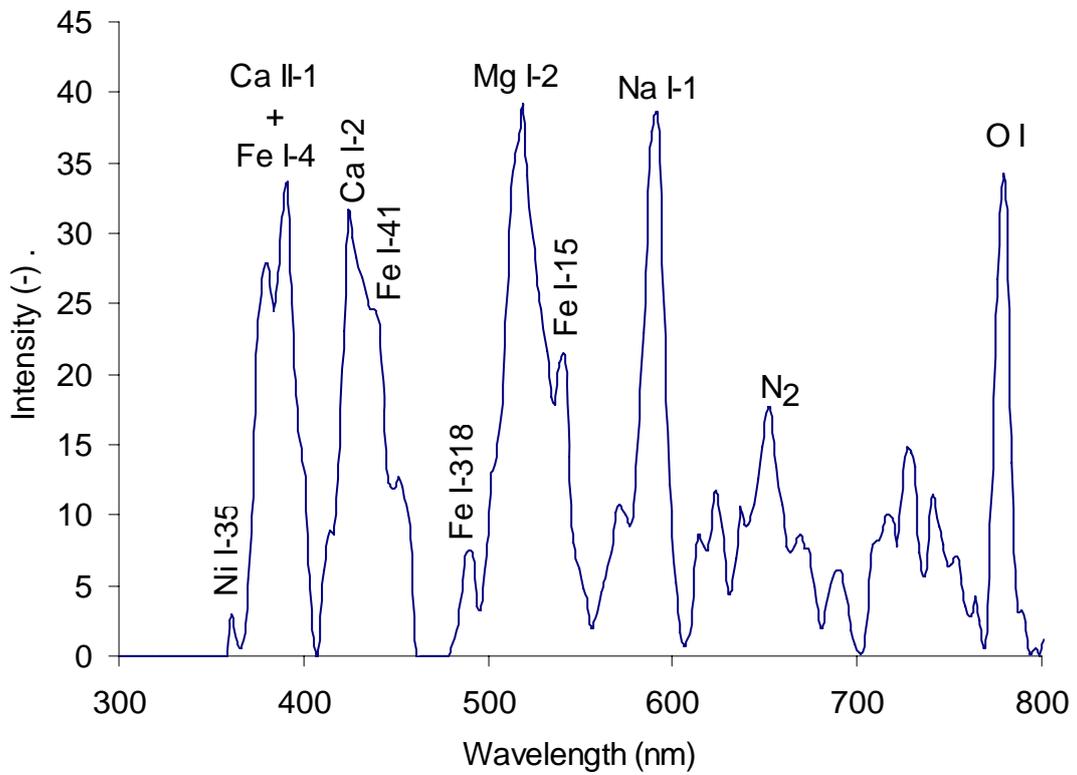

Figure 3. Emission spectrum of the SPMN151209 Geminid bolide. Intensity is expressed in arbitrary units.

13